\documentclass[12pt]{article}
\usepackage[margin=0.95 in]{geometry}
\usepackage{amsmath}
\usepackage{amssymb,amsfonts}
\usepackage[all]{xy}
\usepackage{graphicx}
\usepackage[utf8x]{inputenc}
\usepackage{amsmath}
\usepackage{amssymb}
\usepackage{float}
\usepackage{array}
\usepackage{tikz}
\usepackage{mathtools}
\usepackage{mathrsfs}
\usepackage{hyperref}
\usepackage{cite}
\numberwithin{equation}{section}
\numberwithin{equation}{section}
\numberwithin{table}{section}
\begin{document}

\thispagestyle{empty}

\vspace*{3cm}
{}

\noindent
{\LARGE \bf  Topological Gravity with Non-Compact Matter}
\vskip .4cm
\noindent
\linethickness{.06cm}
\line(10,0){467}
\vskip 1.1cm
\noindent
\noindent
{\large \bf Songyuan Li and Jan Troost}
\vskip 0.25cm
{\em
\noindent
Laboratoire de Physique Th\'eorique
de l'\'Ecole Normale Sup\'erieure \\
\hskip -.05cm
 CNRS,
 PSL Research University and Sorbonne Universit\'e, Paris, France
}
\vskip 1.2cm

\vskip0cm

\noindent {\sc Abstract: } We couple twisted non-compact $N=(2,2)$ supersymmetric  models to topological gravity 
in two dimensions.  We propose expressions for the genus zero correlation functions based on a
Kadomtsev-Petviashvili integrable hierarchy. Moreover, we prove
 recursion relations satisfied by the topological gravity amplitudes at all genera and compute characteristic critical exponents. We discuss the  extent to which moving beyond the $N=2$ central charge barrier opens a window on two-dimensional gravity with central charge larger than one.

\vskip 1cm

\pagebreak

\newpage
\setcounter{tocdepth}{2}
\tableofcontents

\section{Introduction}
Quantum gravity is hard to formulate, let alone integrate in four dimensions. In lower dimensions, gravity is non-dynamical and  therefore easier to study. In particular, the matrix model formulation of two-dimensional gravity was solved three decades ago \cite{Brezin:1990rb,Douglas:1989ve,Gross:1989vs}, and this led to the solution of two-dimensional gravity when coupled to small amounts of matter in the Liouville continuum
 \cite{Knizhnik:1988ak,David:1988hj,Distler:1988jt}
as well as in  the topological formulation \cite{Witten:1988ze,Witten:1989ig}. See e.g. \cite{Ginsparg:1993is,DiFrancesco:1993cyw} for reviews.

In this paper we couple a new form of matter to topological gravity. 

We consider matter topological quantum field theories which arise from $N=2$ non-compact superconformal field theories with central charge $c>3$ after deformation by supersymmetry preserving relevant operators and twisting \cite{Li:2018rcl}.
The chiral ring structure constants, and therefore all correlators of these matter topological quantum field theories were solved for in \cite{Li:2018rcl}. 
In this paper, we first lay bare the integrable structure underlying  the topological quantum field theory.  We  identify the  flows in the Kadomtsev-Petviashvili integrable hierarchy that code the solution of the non-compact matter theory. 

We then exploit the identification of the  integrable hierarchy to propose a generating function for correlation functions after coupling the non-compact topological quantum field theory to gravity, at genus zero.
Using a conformal field theory approach, we furthermore compute the recursion relations satisfied by the topological string amplitudes at all genera, and identify a (half) Virasoro algebra that  governs the perturbative correlation functions. We also comment on the presence of a larger symmetry algebra.

It is known that topological gravity coupled to twisted topological matter reproduces the critical exponents of the minimal unitary models 
with central charge $c<1$ coupled to ordinary two-dimensional gravity \cite{Dijkgraaf:1990qw}. The appropriate supersymmetric quantum field theories to twist are the minimal $N=2$ superconformal field theories at central charge $c=3-6/k_c$ where $k_c$ is a positive integer \cite{Dijkgraaf:1990nc,Li:1990ke,Li:1990fc,Dijkgraaf:1990dj}. 
In the $N=2$ topological formulation, the  central charge barrier is therefore located at $c=3$. {From} the perspective of two-dimensional gravity coupled to ordinary matter, the central charge barrier ($c = 1$) is hard to surmount.
This provides  one motivation for our study of the topological phase of gravity based on an $N=2$ superconformal field theory with central charge $c>3$. We will  discuss to what extent moving beyond the topological central charge barrier gives insight into 
matter with central charge $c>1$ coupled to ordinary two-dimensional gravity.

The plan of our paper includes the identification of the integrable hierarchy for the topological matter model in section \ref{noncompacthierarchy}. In section \ref{noncompacthierarchywithgravity} we propose genus zero 
correlation functions after coupling the model to topological gravity.
Section \ref{recursion} sets up the recursion relation analysis  and identifies a symmetry algebra that will govern the amplitudes at all genera.
We determine scaling behaviors in section \ref{scaling}, using both the approaches
of sections \ref{noncompacthierarchywithgravity} and  \ref{recursion}, and evaluate the result in the broader context
of surmounting the central charge barrier in two-dimensional gravity. We conclude in section \ref{conclusions} with a summary  and  intriguing open problems.

\section{The  Integrable Hierarchy}
\label{noncompacthierarchy}
In this section, we discuss the  Kadomtsev-Petviashvili integrable hierarchy. The hierarchy is realized in a polynomial ring quotiented by the ideal generated by the derivative of a  superpotential \cite{Krichever:1992sw,Dubrovin:1992eu,Dubrovin:1992dz}. To incorporate both the compact and the non-compact $N=2$ superconformal models, we follow the detailed treatment of rational potentials in \cite{Aoyama:1995kv}. We briefly review the integrable hierarchy, and then determine the subset of flows that correspond to the solution of the non-compact topological quantum field theory in \cite{Li:2018rcl}. Our review will be more encompassing since we intend to  exploit the integrable hierarchy to couple the  non-compact model  to topological gravity in section \ref{noncompacthierarchywithgravity}.

\subsection{The Kadomtsev-Petviashvili Integrable Hierarchy}
\label{review}
Our first aim is to embed the solution to the twisted non-compact $N=(2,2)$ supersymmetric model obtained in \cite{Li:2018rcl} in an integrable hierarchy. As in the compact case \cite{Dijkgraaf:1990dj,Krichever:1992sw,Dubrovin:1992eu,Dubrovin:1992dz}, the relevant integrable hierarchy is the dispersionless KP hierarchy \cite{Takasaki:1992ve,Takasaki:1994xh} reduced by the derivative of the superpotential $W$. While the superpotential of the compact models is polynomial, the superpotential of the non-compact $N=(2,2)$ models can be argued to consist of inverse powers of a 
chiral superfield \cite{Ooguri:1995wj,Ashok:2007ui,Li:2018rcl}. If we wish to treat both cases as special cases of a generic model, we are lead to consider rational superpotentials. This generic case was pedagogically exposed in \cite{Aoyama:1995kv} to which we must  refer for many  detailed derivations. Our task here will be to  identify the subset of flows of interest in our physical models, and to treat the peculiarities of the non-compact models with due care. We will reap the reward in section \ref{noncompacthierarchywithgravity}.

A lightning review of \cite{Aoyama:1995kv} starts with the definition of the class of rational superpotentials $W$ under study:
\begin{equation}
W = \frac{X^{k_c}}{k_c} + v_{k_c-2} X^{k_c-2} + \dots
+ v_0 + \frac{v_{-1}}{X-s} + \frac{v_{-2}}{2(X-s)^2} +
\dots + \frac{v_{-k}}{k (X-s)^k} \, . \label{genericsuperpotential}
\end{equation}
The superpotential $W$ has singularities at $X=\infty$ and $X=s$. It has two positive integer parameters $k_c$ and $k$, and $k+k_c$ real coefficients labeled $v_{a}$ and $s$.
We define the Laurent power series $L_c$ and $L$ through the equalities
\begin{equation}
W = \frac{L_c^{k_c}}{k_c} = \frac{L^k}{k} \, , \label{definitionL}
\end{equation}
where $L_c$ is a formal power series expanded at large $X$ and $L$ a formal power series expanded near $X=s$. 
We also define flow generators $Q^{i}$ with integer index $i \in \mathbb{Z}$ in terms of the roots $L_c$ and $L$ of the superpotential $W$ \cite{Aoyama:1995kv}:
\begin{eqnarray}
Q^{i \ge 0} &=& {[} \frac{L_c^{i+1}}{i+1} {]}_{\ge 0}
\nonumber \\
Q^{-1} &=& [ \log L_c ]_{\ge 0} - [ \log L ]_{\le -1}
\nonumber \\
Q^{i \le -2} &=& - [ \frac{L^{|i|-1}}{|i|-1} ]_{\le -1} \, ,
\end{eqnarray}
where the indices on the square bracket indicate which terms in the formal power series to retain.\footnote{E.g. $[ S ]_{\ge 0}$ instructs to keep only the terms with non-negative powers  in the formal power series $S$.}
We introduce an infinite number of coupling constants $t_{i \in \mathbb{Z}}$ 
and demand that the superpotential satisfy the flow equations
\begin{eqnarray}
\partial_{t_i} W = \{ Q^i , W \} = \partial_X Q^i \partial_{t_0} W - \partial_{t_0} Q^i \partial_X W \, . \label{zeroformulation}
\end{eqnarray}
These equations define the dispersionless  Kadomtsev-Petviashvili hierarchy reduced by the derivative of the superpotential $W$. The equations are mutually compatible because the connection $Q$  has zero curvature. The flows along times $t_i$ commute and the system is integrable \cite{Aoyama:1995kv}. 
The infinite set of operators that we wish to study are given by the derivatives of the generators $Q^i$:
\begin{equation}
\phi^i = \partial_X Q^i \, . \label{phis}
\end{equation}
The operators $\phi^i$ are elements of the polynomial ring $\mathbb{C}[X,(X-s)^{-1}]$. If we divide the ring by the ideal generated by the derivative  of the superpotential $\partial_X W=
\phi^{k_c-1}-\phi^{-k-1}$, then the quotient ring has a  basis of operators  $\phi^\alpha$ where $\alpha$ takes values in the set $\Delta=\{ -k, -k+1, \dots, k_c-1 \}$. We refer to the operators in this basis as primaries.
Crucially, it is also possible to reformulate the KP hierarchy democratically with respect to any other time $t_j$ and such that the generators $Q^i$ satisfy
\begin{equation}
\partial_{X} Q^i \partial_{t_j} W -
\partial_{X} W \partial_{t_j} Q^i
=
\partial_{X} Q^j \partial_{t_i} W -
\partial_{X} W \partial_{t_i} Q^j \, .
\end{equation}
The original formulation (\ref{zeroformulation}) corresponds to the choice $j=0$. If one instead singles out $t_j=t_{\alpha_0}$ where $\alpha_0 \in \Delta$, then a new set of primary fields is given by $\tilde{\phi}^\alpha = \phi^\alpha/\phi^{\alpha_0}$ modulo $W'$.
The identity element becomes $\tilde{\phi}^{\alpha_0}$. The solutions of the hierarchy remain the same.

It is useful to introduce the universal coordinates $u^{i \in \mathbb{Z}}$ by inverting the formal series $L$ and $L_c$, 
and expand the formal variable $X$ in terms of $L_c$, at large $L_c$, or in terms of $L$ at large $L$:
\begin{eqnarray}
X &=& L_c- \frac{u^0}{L_c}-\dots- \frac{u^i}{L_c^{i+1}} + \dots
\nonumber \\
X &=& u^{-1} + \frac{u^{-2}}{L} + \dots + \frac{u^{-i}}{L^{i-1}} +\dots 
\end{eqnarray}
The universal coordinates $u^i$ are then residues of the roots of the superpotential:
\begin{eqnarray}
u^{i \ge 0} &=& Res_{\infty} \frac{L_c^{i+1}}{i+1}
\nonumber \\
u^{-1} &=& Res_{\infty} \log L_c + Res_{s} \log L
\nonumber \\
u^{i \le -2} &=& Res_{s} \frac{L^{|i|-1}}{|i|-1} \, .   \label{uresidues}
\end{eqnarray}
The derivative of the superpotential with respect to the primary universal coordinates $u^\alpha$ gives the primary fields $\phi^\beta$, multiplied by a metric $\eta_{\alpha \beta}$:
\begin{equation}
\partial_{u^\alpha} W = \phi_\alpha = \eta_{\alpha \beta} \phi^\beta \, ,  \label{definitionprimaries}
\end{equation}
where the metric $\eta_{ \alpha \beta}$ equals
\begin{eqnarray}
\eta_{\alpha \beta} &=& \delta_{\alpha+\beta,k_c-2}
\quad \, \mbox{for} \quad \alpha,\beta \ge -1
\nonumber \\
\eta_{\alpha \beta} &=& \delta_{\alpha+ \beta,-k-2}
\quad \mbox{for} \quad \alpha,\beta \le -2 \, , \label{metric}
\end{eqnarray}
and zero otherwise. Moreover, the coefficients $v_a$ of the superpotential $W$ (\ref{genericsuperpotential}) can be computed in terms of the universal coordinates $u_\alpha$ as \cite{Aoyama:1995kv}
\begin{eqnarray}
v_{-1 \le -a \le -k} &=& \sum_{\substack{\alpha_1 + \dots + \alpha_a = (a-1)k+a \\ \alpha_i > 0}} u_{-\alpha_1} u_{-\alpha_2}  \dots u_{-\alpha_a}
\nonumber \\
v_{0 \le a \le k_c-2} &=& u_a + \sum_{b=2}^{k_c-2-a} \frac{(a+b-1)!}{a ! b!} \sum_{\substack{\alpha_1+\dots+\alpha_b=(b-1)(k_c-1)+a+b-1 \\ 0 \le \alpha_i \le k_c-2}} u_{\alpha_1} \dots u_{\alpha_b}
\nonumber \\
s &=& u_{k_c-1} \, .
\end{eqnarray}
One can also introduce Gelfand-Dickey potentials $G^{ij}$ which satisfy
\begin{equation}
\partial_{t_j} u^i = \partial_{t_0} G^{ij} \, ,
\end{equation}
and they can be explicitly computed  in terms of the roots $L$ and $L_c$ and the fields $\phi^i$ \cite{Aoyama:1995kv}
\begin{eqnarray}
G^{ij} &=& \frac{1}{i+1} Res_\infty L_c^{i+1} \phi^j \quad \mbox{for} \quad i,j \ge 0
\nonumber \\
G^{-i,-j} &=& \frac{1}{i-1} Res_{s} L^{i-1} \phi^{-j} \quad \mbox{for} \quad  i \ge 2, \quad j \ge 1 
\nonumber \\
G^{i,-j} &=& \frac{1}{i+1} Res_{\infty} L_c^{i+1} \phi^{-j} \quad \mbox{for} \quad  i \ge 0, \quad j \ge 1 
\nonumber \\
G^{-1,-1} &=& Res_\infty \log L_c \phi^{-1} + Res_s \log L \phi^{-1} = \log u_{-k} \, .
\label{twopointfunctions}
\end{eqnarray}
These potentials are symmetric in their indices. While we have $u^i=G^{i0}$, we can also define generalized universal coordinates
$\tilde{u}^i=G^{i \alpha_0}$ which are related to the generalized primary fields $\tilde{\phi}^\alpha$ in the same manner as the universal coordinates are related to the primaries:
\begin{equation}
\partial_{\tilde{u}^\alpha} W  = \tilde{\phi}_\alpha \, .
\end{equation}
An important result is that in the small phase spanned by the primaries 
with index in $\Delta$, the KP hierarchy has a solution for 
each $\alpha_0 \in \Delta$ given by \cite{Aoyama:1995kv}
\begin{equation}
\tilde{u}^\alpha = \eta^{\alpha \beta} t_\beta \, . \label{smallphasespacesolution}
\end{equation}
For the small phase space solution (\ref{smallphasespacesolution}), the (generalized) two-point function $G^{\alpha \beta}$ can be  integrated twice with respect to the universal coordinates $u_\alpha$
to  a generating function $F_m$. 
\begin{equation}
G^{\alpha \beta} = \partial_{u_\alpha} \partial_{u_\beta} F_m(u) \, .
\end{equation}
One can define (generalized) $N$-point functions as $N$-fold derivatives of the generating function $F_m$ of primary (matter) correlation functions and obtain residue formulas for these $N$-point functions. We again refer the reader to \cite{Aoyama:1995kv} for more and more explicit formulas. We have gathered ample data to excavate the non-compact topological quantum field theory correlators from these formulas.

\subsection{The  Non-Compact Topological Theory}
\label{limit}
\label{TQFT}
In this subsection, we wish to show how to reduce the KP integrable hierarchy reviewed in subsection \ref{review}
to the solution of the non-compact topological quantum field theory of \cite{Li:2018rcl}. The topological theory
is the twisted $N=2$ Liouville theory at 
central charge $c=3+6/k$ and with asymptotic radius $\sqrt{k \alpha'}$, deformed by relevant operators. The number of strictly normalizable ground states in the Ramond-Ramond sector of the theory equals $k-1$, and this is the dimension of the chiral ring. At the critical point (i.e. the topological conformal field theory), the ring is defined by the polynomials in $X^{-2} \mathbb{C} [ X^{-1}]$, where 
we consider the element $X^{-k-1}$ to be trivial. The space of these topological quantum field theories is spanned by the deformations of the quotient ring. It can succinctly be  summarized in terms of coupling constants $c_{\alpha \beta \gamma}(t_\alpha)$ which are functions of  parameters $t_\alpha$. The structure constants determine the correlation functions of the topological quantum field theory on all Riemann surfaces.
The explicit solution of the theory was obtained in \cite{Li:2018rcl}   with  combinatorial means. Here, we embed the solution in the KP integrable hierarchy.

Before we do so we remind the reader of the fact that it is straightforward to embed the twisted compact $N=2$ model into the KP integrable hierarchy \cite{Dijkgraaf:1990dj}. The deformed minimal superconformal model at central charge $c=3-6/k_c$ is solved by the KP integrable hierarchy we reviewed, where we identify $k_c$ in the integrable hierarchy with the level $k_c$ of the superconformal model and where we restrict to $k=0$.
The  small phase space solution (\ref{smallphasespacesolution}) for $\alpha_0=0$ then entirely agrees with the solution of the topological quantum field theory
\cite{Dijkgraaf:1990dj} -- it is easy to verify this by matching explicit formulas. The identification of the non-compact model requires a little  extra work.

The superpotential we wish to use is \cite{Li:2018rcl} 
\begin{equation}
W = \sum_{n=2}^k \frac{v_{-n}}{n} X^{-n} \, .
\label{strictW}
\end{equation}
In order to obtain this superpotential form the generic case (\ref{genericsuperpotential}),  we propose to take $k_c=1$, since this is the minimal value allowed in the approach of \cite{Aoyama:1995kv}. The first remark  is that we can then trade the pole $s$ for a constant term $s$ by shifting the formal variable $X$.  We also have to deal with the fact that the leading term in the superpotential (\ref{genericsuperpotential}) has coefficient one for
the leading operator $X$ (at $k_c=1$). Thus, we must further reduce. In order to do so,  we rescale
\begin{eqnarray}
 X & \rightarrow & \epsilon X
 \nonumber \\
 u_{-\alpha \le -1} & \rightarrow & \epsilon u_{-\alpha \le -1} \, ,
 \end{eqnarray}
 and then take $\epsilon \rightarrow 0$. This leaves us with a model where the leading term is a constant. The last step in the reduction
 is that we set
 $u_0=0=u_{-1}$ in order to eliminate both the constant and the $X^{-1}$ term, and match the desired superpotential (\ref{strictW}).  At the same time, we restrict the spectrum of the model to the set of primaries
$\Delta^{strict}=\{-2,-3, \dots ,-k\} \subset \Delta$.
 It is then straightforward  to check that the definition of
 the root $L$ (\ref{definitionL}), the operators $\phi^\alpha$ (\ref{definitionprimaries}), and the universal coordinates
 $u^\alpha$ (\ref{uresidues}) when restricted to the small phase space solution (\ref{smallphasespacesolution}) match those of the solution provided in 
 \cite{Li:2018rcl}. 
 
Finally, for future purposes, we note the following properties. Firstly, the operators with labels $\alpha \ge -1$ have a metric $\eta_{\alpha \beta}$ (see
 (\ref{metric})) defined in terms of the compact level $k_c$. These include the identity operator $\phi^0$ and
 the operator $\phi^{-1}$. These two operators  become conjugate at $k_c=1$. The primary operators with labels $\alpha \le -2$ on the other hand have a metric $\eta_{\alpha \beta}$ set by the non-compact level $k$. The latter fact  confirms the identification of the topological metric $\eta_{\alpha \beta}$ argued for on different grounds in \cite{Li:2018rcl}. 
Moreover,  the primary operators in the  $k_c=1$ model before taking the limit are
\begin{eqnarray}
\phi^0 &=& 1
\nonumber \\
\phi^{-1} &=& X^{-1}
\nonumber \\
\phi^{i \le -2} &=& [\phi^{i \le -2} ]_{\le -2} \, .
\label{operators}
\end{eqnarray}
These equations show that the latter operators do not mix with the former.

In this manner, we have embedded the combinatorial solution of \cite{Li:2018rcl} into the dispersionless KP integrable hierarchy of \cite{Aoyama:1995kv}.  Thus, the model of \cite{Li:2018rcl} is integrable.
In appendix \ref{TQFTexample} we provide an example solution at $k=3$, and give more details on how the limiting procedure described above takes us from the  KP hierarchy \cite{Aoyama:1995kv} into the  solution presented in \cite{Li:2018rcl}.

\section{Topological Gravity with Non-Compact Matter}
\label{noncompacthierarchywithgravity}
In section \ref{noncompacthierarchy} we  reviewed the dispersionless KP hierarchy  and explained how the matter non-compact topological quantum field theory fits into the hierarchy. In this manner, we have gained considerable control over the integrable extensions of the topological quantum field theory. Indeed, the KP hierarchy allows for a larger set of times and operators, all compatible with integrability. There are two extensions in particular that are of interest to us. The extension by the identity operator and the time $t_0$, as well as the extension by times coupling to the gravitational descendants of the matter primary operators. In this section, we concentrate on the latter, under the assumption that 
 the non-compact topological quantum field theory coupled to topological gravity in two dimensions is still governed by the KP integrable hierarchy. In this manner, we  obtain  the  correlation functions of
primaries and gravitational descendants  on genus zero Riemann surfaces (i.e. at topological string tree level).

The extension proceeds as follows. The infinite set of fields $\phi^{i \le -2}$  are related by the recursion relation
\begin{equation}
\phi^{-Nk+\alpha} = \frac{1}{Nk-\alpha-1} \partial_{u_0} \phi^{-(N+1)k+\alpha} \, , 
\end{equation}
for $N \ge 0$ and  $\alpha \le -2$.\footnote{There are analogous formulas for 
$\alpha \ge -1$ \cite{Aoyama:1995kv}. }
The gravitational descendants  are related to these infinite
towers built on primaries.
For convenience, we define the gravitational descendants to have a different normalization. Using the constants $d_{N,\alpha}$
\begin{equation}
d_{N,\alpha}=\Big( (|\alpha|-1)(|\alpha|-1+k)\dots (|\alpha|-1+(N-1)k) \Big)^{-1} \, ,
\end{equation}
we define the gravitational descendants $\sigma_N(\phi^\alpha)$ of the matter primaries $\phi^{\alpha \le -2}$ as
\begin{equation}
\sigma_N(\phi^\alpha) = d_{N,\alpha} \phi^{-Nk+ \alpha} \, .
\end{equation}
We thus find the recursion relation among gravitational descendants
\begin{equation}
\sigma_{N-1}(\phi^\alpha) = \partial_{u_0} \sigma_N (\phi^\alpha) \, ,
\end{equation}
where we remind the reader that equations hold modulo the derivative $W'$ of the superpotential.
The gravitational times are defined with a related change in normalization
\begin{equation}
t_{N,\alpha} = d^{-1}_{N,\alpha} t_{-Nk+\alpha} \, .
\end{equation}
All these definitions can be extended to all matter primaries in the 
spectrum $\Delta$ \cite{Aoyama:1995kv}.

The relation to topological gravity becomes manifest by observing that all the gravitational descendants $\sigma_N(\phi^\alpha)$ satisfy 
 the topological recursion relation \cite{Witten:1989ig,Dijkgraaf:1990nc}
\begin{equation}
\langle \sigma_N(\phi^\alpha) AB \rangle =
\langle \sigma_{N-1} (\phi^\alpha) \phi^\beta \rangle 
\langle \phi_\beta AB \rangle \, ,
\label{topologicalrecursion}
\end{equation}
where the sum is over a complete set if primaries with labels in $\Delta$.
The proof of the topological recursion (\ref{topologicalrecursion}) passes through the decomposition of the descendants into primaries \cite{Aoyama:1995kv}\footnote{We caution the reader that in this section, and in appendix \ref{examples}, the two-point functions should be read as second derivatives
of the generating function of correlation functions. The physical two-point functions are obtained by a further derivative with respect to the coordinate that couples to the identity operator \cite{Aoyama:1995kv}.}
\begin{equation}
\sigma_N(\phi^\alpha) = \langle \sigma_{N-1}(\phi^\alpha) \phi^\beta \rangle \phi_\beta \, ,
\end{equation}
where the sum is over the set of primaries $\Delta$. Importantly, 
for the model with $k_c=1$,
we note that the descendants of the primaries in the range
$\alpha \le -2$ can be expressed in terms of the primaries in the range 
$\alpha \le -2$. This follows from the separation of operators observed in equations (\ref{operators}) that persists for the gravitational descendants in the limit theory described in section \ref{noncompacthierarchy}. Indeed,
the operators $\phi^{i \le -2}$ only contain terms $X^j$ with $j \le -2$, as does the derivative of the superpotential
in the limit theory.  Thus, when we restrict the 
primaries to the range $\Delta^{strict}=
\{ -2, -3 , \dots , -k \} \subset \Delta$,
 the topological recursion  and the descendant decomposition remain valid.   This is crucial in order for  the non-compact theory with spectrum $\Delta^{strict}$ to be consistent.
 
 The main property that allows to solve for all correlation functions of the non-compact model to gravity, is the fact that the
  generalized two-point functions $G^{ij}$ (\ref{twopointfunctions}):
 \begin{equation}
 G^{ij} (u) = \langle \phi^i  \phi^j \rangle \,  , \label{constitutive}
 \end{equation}
 as a function of the universal coordinates $u$ remain invariant after coupling to gravity.
Thus, the equations (\ref{constitutive}) are known as  constitutive equations  \cite{Dijkgraaf:1990nc}.  Only the relation between
the universal coordinates and the  primary couplings $t_{0,\alpha}$ undergoes a renormalization. Indeed, the explicit  solution of the full KP hierarchy including the infinite set of times $t_{N,\alpha}$ is given in terms of universal coordinates dependent on all these times, as follows  \cite{Aoyama:1995kv}
\begin{eqnarray}
u^\alpha(t_{j \in \mathbb{Z}})=\hat{u}^\alpha(\hat{t}_\beta)
\end{eqnarray}
where the $\hat{u}$ are determined by the solution on the small phase space
\begin{equation}
\hat{u}^\alpha(\hat{t}_\gamma) = u^\alpha(\dots,0,\hat{t}_{-k},\hat{t}_{-k+1},\dots,\hat{t}_{k_c-1},0,\dots)
\end{equation}
and the renormalized primary times $\hat{t}_\beta$ are
\begin{equation}
\nonumber \\
\hat{t}_\beta = t_\beta + \sum_{\alpha \in \Delta, N\ge 1}
\langle \sigma_{N-1}(\phi^\alpha) \phi_\beta \rangle t_{N,\alpha} \, .
\end{equation}
Thus, the full solution is determined by the small phase space solution and the (generalized) two-point functions of descendants with primaries. This is entirely consistent with the fact that the gravitational descendants can
be decomposed into matter primaries, such that coupling the descendants 
renormalizes the primary times also from the viewpoint of a renormalized action (rewritten in terms of primaries only). 
We note  that descendants in the restricted spectrum only renormalize times in the restricted spectrum.

The explicit generator $F_0$ of genus zero correlation functions is determined by integrating the two-point function $G^{ij}(u)$ twice
with respect to times, and by renormalization. The explicit formula  \cite{Aoyama:1995kv}
\begin{equation}
F_0(t_{N,\alpha}) = \frac{1}{2} \sum_{\alpha,\beta \in \Delta, N,M \ge 0}
\tilde{t}_{N,\alpha} \tilde{t}_{M,\beta} \langle \sigma_{N}(\phi^\alpha) \sigma_M (\phi^\beta) \rangle (t) \, ,
\label{F}
\end{equation}
is
obtained from topological recursion which implies homogeneity
in the shifted time variables $\tilde{t}_\beta$
\begin{equation}
\tilde{t}_\beta = {t}_\beta + C_{N,\beta} \quad \mbox{for} \quad N \ge 1 \, .
\end{equation}
These times  satisfy the equation
\begin{equation}
\tilde{t}_\alpha + \sum_{\beta,N \ge 1} \langle \sigma_{N-1} (\phi^\beta) \phi_\alpha \rangle \tilde{t}_\beta = 0
\end{equation}
and the constants $C_{N,\beta}$ obey
\begin{equation}
\hat{t}_\alpha = - \sum_{\beta,N \ge 1} \langle \sigma_{N-1} (\phi^\beta) \phi_\alpha \rangle C_{N,\beta} \, .
\end{equation}
Finally, we propose that the renormalized solution to the KP hierarchy (\ref{F}) generates the genus zero correlation functions for topological gravity coupled to topologically twisted and deformed non-compact $N=2$ superconformal field
theories with central charge $c=3+\frac{6}{k}>3$ after applying the limiting procedure described in section \ref{noncompacthierarchy} to the theory. The limiting procedure is consistent, also after coupling to gravity.

To illustrate the proposal, we provide explicit formulas for the generating function $F_0$ for a non-compact model at level $k=3$ with a non-zero gravitational coupling in appendix \ref{gravitationalexample}.

\section{The Recursion Relations}
\label{recursion}
Topological gravity amplitudes satisfy Virasoro or W-algebra constraints that determine amplitudes recursively, up to initialization \cite{Witten:1989ig,Dijkgraaf:1990nc,Verlinde:1990ku,Dijkgraaf:1990rs,Fukuma:1990jw}.
In this section, we closely follow the conformal field theory derivation of these constraints
pioneered in \cite{Verlinde:1990ku} and applied to twisted $N=2$ minimal models in \cite{Li:1990ke,Li:1990fc}. We generalize the calculation to other matter spectra, which include the non-compact models of interest in this paper.\footnote{
A seemingly similar analysis was performed in \cite{Yu:1992yp}. While there is technical overlap, our analysis differs in important ways, for instance  in the identification of the  spectrum of the non-compact matter theory.}

\subsection{The Derivation of the Recursion Relations}
We follow \cite{Verlinde:1990ku,Li:1990fc} closely and must refer to these references for background and many of the underlying details. In particular, the conformal field theory approach to topological gravity is neatly exhibited in \cite{Verlinde:1990ku}, and the further coupling to matter is detailed in \cite{Li:1990fc}. Our task is to carefully identify the central charge dependence of the calculations in these papers, as well as a conceptual change related to the matter spectrum.
We recall that the gravitational descendants are often represented by the conformal field theory vertex operators\footnote{For $
\alpha \le -2 $, we expect these vertex operators
to be closely related to the operators $\sigma_N(\phi^\alpha)$ introduced in section \ref{noncompacthierarchywithgravity}.}
\begin{equation}
\sigma_{n,\alpha}' = O^m_\alpha P \gamma_0^n  \, ,
\end{equation}
where  $O^m_\alpha$ is a matter primary before coupling to gravity, $P$ is the puncture operator and $\gamma_0$ is the commuting ghost superpartner of the fermionic ghost associated to local Lorentz transformations \cite{Witten:1989ig,Verlinde:1990ku}.
The selection rule on non-zero amplitudes depends on the matter central charge $c$ as well as the matter R-charges $q_\alpha$.
The ghost number conservation rule generically reads \cite{Verlinde:1990ku,Li:1990fc}
\begin{equation}
\sum_i (n_i -1 + q_{\alpha_i}) = (g-1)(3-\frac{c}{3}) \, , 
\label{ghostnumberconservation}
\end{equation}
where we assumed that we are dealing with two-form vertex operators to be integrated over the world sheet of genus $g$ enumerated by the index  $i$. For each insertion $i$, we have a descendant of order $n_i$ and R-charge $q_{\alpha_i}$.
The computational technique of \cite{Verlinde:1990ku} consists in distributing curvature singularities on all the vertex operators in the amplitude evenly, and in such a way that they add up to the Euler
number $2g-2$, thus reducing the calculation of all gravitational scattering amplitudes to contact interactions. In order to implement this, we can add a curvature charge $c_{n,\alpha}$ to each
vertex operator $\sigma'_{n,\alpha}$ equal to 
\begin{equation}
c_{n,\alpha}=\frac{6( n-1+q_\alpha)}{9-c} \, , \label{curvatureconstants}
\end{equation}
and we denote
the new vertex operators by $\sigma_{n,\alpha}$. Then, by the selection rule (\ref{ghostnumberconservation}), the curvature contributions add up to  $2g-2$, as required.

We expect a Virasoro algebra to be associated to the action of the puncture operator and its descendants on the correlation functions \cite{Verlinde:1990ku,Li:1990fc}. We must stress that the identity operator is  not necessarily part of the spectrum of the theory, and that therefore, the puncture operator and its descendants may also be absent from the spectrum. In this case, we will view these operators as {\em acting} on the collection of all physical amplitudes. 
This is an important conceptual shift. We will use these operators as a ladder to be climbed, then to be discarded.

Still, we expect these generalized amplitudes are governed by contact terms only \cite{Verlinde:1990ku}. 
The action of the puncture descendants $\sigma_{m,0}=\sigma_m$ gives rise to a contact term with all other operator insertions, with coefficient $c_{n,\alpha}+1$, determined by the curvature attached to the operator.  There are however further contributions when the Riemann surface degenerates into a surface of one genus less (by collapsing a handle) or by
splitting into two separate surfaces whose genera add up to the genus of the original surface \cite{Witten:1989ig}.
After careful examination of the degeneration limits (e.g. in a conformal field theory language), the total recursion relation for a genus $g$ amplitude is computed to be of the form \cite{Verlinde:1990ku,Li:1990fc}
\begin{align}
\langle \sigma_m \prod_{i \in S} \sigma_{n_i,\alpha_i} \rangle_g
=& \sum_{i \in S} (c_{n_i,\alpha_i}+1) \langle 
\sigma_{m+n_i-1,\alpha_i}
\prod_{j \neq i} \sigma_{n_i,\alpha_i}\rangle_{g} \nonumber 
\\
&+
 \sum_{\alpha, \beta \in \Delta_m} \sum_{n=2}^m
b_\alpha \eta^{\alpha \beta}
\langle \sigma_{n-2,\alpha} \sigma_{m-n,\beta}
 \prod_{i \in S} \sigma_{n_i,\alpha_i} 
 \rangle_{g-1} \nonumber
\\
&+
\sum_{\alpha,\beta \in \Delta_m} \sum_{n=2}^m
\sum_{X \cup Y = S}
b'_\alpha \eta^{\alpha \beta}
 \langle 
\sigma_{n,\alpha} \prod_{i \in X} \sigma_{n_i,\alpha_i}
\rangle_{g_X} \langle
\sigma_{m-n,\beta} \prod_{j \in Y} \sigma_{n_j,\alpha_j}
\rangle_{g-g_X} \, ,
\label{recursionrelation}
\end{align}
where $S$ is the set of insertions, $\Delta_m$ captures the matter spectrum, $\eta_{\alpha \beta}$ is a normalized
matter metric, and $b_\alpha$ and $b'_\alpha$ are constants that we wish to determine.
We can then compute the correlation functions
 $\langle \sigma_2 P \rangle_1$ as well as $\langle \sigma_2 P O_{\alpha_1} O_{\alpha_2} O_{\alpha_3} \rangle_0$ using the recursion relation,
 or alternatively, the puncture equation \cite{Witten:1989ig}.
 This fixes the constants $b_\alpha$ and $b'_\alpha$  in terms of the normalization $A_\alpha$ of the gravitationally dressed primary operators $O_\alpha$,
where $\langle P O_\alpha O_\beta \rangle_0=A_\alpha \eta_{\alpha \beta}$, and in terms of the genus one one-point function $\langle \sigma_1 \rangle_1$ \cite{Verlinde:1990ku,Li:1990fc}. The detailed calculations involve the steps
\begin{eqnarray}
\langle \sigma_2 P \rangle_1 &=& (c_{2,0}+1) \langle \sigma_1 \rangle_1
\nonumber \\
&=& (c_{0,0}+1) \langle \sigma_{1} \rangle_1 + \sum_\alpha b_\alpha \eta^{\alpha \beta} \langle O_\alpha O_\beta P \rangle_0
= (c_{0,0}+1) \langle \sigma_{1} \rangle_1 + \sum_{\alpha} b_\alpha A_\alpha
\end{eqnarray}
as well as
\begin{eqnarray}
\langle \sigma_2 P O_{\alpha_1} O_{\alpha_2} O_{\alpha_3} \rangle_0
&=& (c_{2,0}+1) \langle \sigma_1 O_{\alpha_1} O_{\alpha_2} O_{\alpha_3} \rangle_0 = (c_{2,0}+1) \langle  O_{\alpha_1} O_{\alpha_2} O_{\alpha_3} \rangle_0
\nonumber \\
&=& (c_{0,0}+1) \langle \sigma_1 O_{\alpha_1} O_{\alpha_2} O_{\alpha_3} \rangle_0
+ \sum_{i=1}^3 (c_{1,\alpha_i}+1) \langle \sigma_{1,\alpha_i} P \prod_{j \neq i} O_{\alpha_j} \rangle_0
\nonumber \\
& & +
\sum_{\alpha,\beta \in \Delta_m} 2 b'_\alpha \eta^{\alpha \beta} \sum_{{ i=1} 
}^3 \langle O_\alpha P O_{\alpha_i} \rangle_0
\langle O_{\beta} \prod_{j \neq i} O_{\alpha_j}  \rangle_0
\nonumber \\
&=& (c_{0,0}+1) \langle O_{\alpha_1} O_{\alpha_2} O_{\alpha_3} \rangle_0 +
\sum_{i=1}^3 (c_{1,\alpha_i}+1) (c_{0,\alpha_i}+1) \langle O_{\alpha_1}  O_{\alpha_2} O_{\alpha_3} \rangle_0
\nonumber \\
& & +
\sum_{i=1}^3 2 b'_{\alpha_i}  A_{\alpha_i} 
\langle O_{\alpha_1}  O_{\alpha_2} O_{\alpha_3}  \rangle_0 \, ,
\end{eqnarray}
where we used that $\sum_{i=1}^3 (c_{1,\alpha_i}+1) = 1$ from ghost number conservation. We applied either  the puncture equation  (in the first line) or the recursion formula for $\sigma_2$ first. We also assumed that the metric $\eta^{\alpha \beta}$ is either zero or one, and couples a given primary to another (or the same) unique primary. We use that for $q_\alpha + q_\beta = c/3$, we have $b'_\alpha=b'_\beta$ and $A_\alpha=A_\beta$. After plugging in the known curvature constants (\ref{curvatureconstants}), we then find
{from} the five-point function that
\begin{equation}
b'_\alpha A_\alpha = \frac{18}{(c-9)^2} (q_\alpha-\frac{c}{6}+\frac{1}{2})(\frac{c}{6}-q_\alpha+\frac{1}{2}) 
\, ,
\label{bprime}
\end{equation}
and from the two-point function we conclude
\begin{equation}
 \sum_{\alpha \in \Delta_m} b_\alpha  A_\alpha =
\frac{12}{9-c} \langle \sigma_1 \rangle_1   \, .
\label{consistency}
\end{equation}
Furthermore, consistency requires that the constants $b_\alpha$ and $b'_\alpha$ are proportional \cite{Verlinde:1990ku,Li:1990fc} 
\begin{equation}
b_\alpha = b'_\alpha a \, ,
\end{equation}
and we obtain that the proportionality factor $a$ equals
\begin{equation}
a = \frac{12}{9-c} \,  \frac{\langle \sigma_1 \rangle_1}{    \sum_{\alpha \in \Delta_m} b_\alpha' A_\alpha} \, .
\end{equation}
Thus, the matter central charge $c$, the  spectrum $\Delta_m$, the  R-charges $q_\alpha$,  and the toroidal 
one-point function determine the constants $b_\alpha, b'_\alpha$ in the recursion relation (\ref{recursionrelation}).
Both constants are proportional to the inverse metric normalization constant $A_\alpha^{-1}$. This makes it clear that the 
recursion relations are normalization independent.
Note also that the  metric $A_\alpha \eta_{\alpha \beta}$ is a genus zero amplitude and therefore depends on the string coupling constant. 
Finally, it is natural to suppose that the degeneration limits of pinching either a topologically trivial or non-trivial cycle are locally equivalent
processes, and this leads one to expect that $a=1$, independently of the theory at hand. One can check that for instance for pure gravity,
 this is implied by the known result $\langle \sigma_{1} \rangle_1=1/24$.
We will suppose $a=1$ to be true from now on.

Let us compute the implications of these results more explicitly still in two cases. Firstly, for the $N=2$ minimal model at central charge $c=3-6/k_c$, we perform the following summation:
\begin{equation}
\sum_{\alpha \in \Delta_m} (q_\alpha-\frac{c}{6}+\frac{1}{2})(\frac{c}{6}-q_\alpha+\frac{1}{2}) = \frac{(k_c-1)(k_c+1)}{6k_c}  \, .
\label{chargesum}
\end{equation}
This is consistent with the formulas $a=1$ and $\langle \sigma_1 \rangle_1 = (k_c-1)/24$ \cite{Li:1990fc}.\footnote{We read the latter formula as a result of summing over two-point functions of gravitationally dressed operators. See equation (\ref{chargesum}).}
For the non-compact $N=2$ theory with central charge $c=3+6/k$, the charge spectrum is the same \cite{Li:2018rcl} and the sum (\ref{chargesum}) is therefore the
same as well (with the replacement $k_c \rightarrow k$). 
Assuming $a=1$ as we do, we find that the toroidal one-point function in the non-compact theory equals $\langle \sigma_1 \rangle_1 = (k+1)/24$.

\subsection{The Differential Equations}
In this subsection, we wish to recast the recursion relations (\ref{recursionrelation}) in terms of differential equations satisfied by a generator of correlation functions.
We define the generator of all genus connected correlation functions $F$ including the coupling to the
identity and its descendants. If we wish to restrict the matter spectrum, we view the extra operators 
as operators acting on the physical correlators. We define \cite{Dijkgraaf:1990qw}
\begin{equation}
F=\log Z = \langle \exp (t_{n,\alpha} \int \sigma_{n,\alpha}) \rangle
= \sum_{g \ge 0} \lambda^{2g-2} \langle \exp (t_{n,\alpha} \int \sigma_{n,\alpha}) \rangle_g = \sum_{g \ge 0} \lambda^{2g-2} F_g \, ,
\end{equation}
where we include the operators $\alpha=0$ in the sum. The parameter $\lambda$ corresponds to a genus counting parameter or string coupling constant,
and $t_{n,\alpha}$ are primary and gravitational descendants couplings.
All genus correlation functions can be obtained from the derivatives of the generator $F$ 
\begin{equation}
\langle \prod \sigma_{n_i,\alpha_i} \rangle = \prod_i \partial_{n_i,\alpha_i} \log Z = \sum_{g \ge 0} \lambda^{2g-2} \langle \prod \sigma_{n_i,\alpha_i} \rangle_g \, .
\end{equation}
When we sum  the recursion relation (\ref{recursionrelation}) over genera weighted by $\lambda^2$ we find
\begin{align}
    \langle \sigma_{m,0}\prod_{i=1}^{s}\sigma_{n_i,\alpha_i} \rangle = & \sum_{i=1}^s (c_{n_i,\alpha_i}+1) \langle \sigma_{m+n_i-1} \prod_{j \neq i} \sigma_{n_j,\alpha_j} \rangle \nonumber  \\
 & + \lambda^2 \sum_{n=2}^m \sum_{\alpha,\beta \in {\Delta}_m}  \Big( b_\alpha \langle \sigma_{n-2,\alpha} \eta^{\alpha\beta} \sigma_{m-n,\beta} \prod_{i=1}^s \sigma_{n_i,\alpha_i} \rangle \nonumber \\ &
\quad \quad  \quad \quad \quad + b^\prime_\alpha \sum_{S=X \cup Y} \langle \sigma_{n-2,\alpha} \prod_{i \in X} \sigma_{n_i,\alpha_i} \rangle \eta^{\alpha\beta}  \langle \sigma_{m-n,\beta} \prod_{j \in Y} \sigma_{n_j,\alpha_j} \rangle \Big) \, .
\end{align}
The expansion of $Z=e^F$ both in the string coupling constant  and in the perturbation parameters $t_{n,\alpha}$ is
strongly constrained by the recursion relation. The universality of the degeneration of Riemann
surfaces (or the property that $a=1$  -- see above) is tied to the fact that we can summarize the terms arising from degenerating surfaces in terms
of a differential operator acting on the function $Z$.
The differential constraint on the generator $Z=e^F$ for $m \ge 2$ is equal to
\begin{equation}
\partial_{t_m} Z = \sum_{n=0}^\infty \sum_{\alpha \in \Delta_m \cup \{ 0 \}} (c_{n,\alpha}+1) t_{n,\alpha} \partial_{t_{n+m-1,\alpha}}Z + \lambda^2 \sum_{\alpha,\beta \in {\Delta }_m} \sum_{n=2}^m b_\alpha \eta^{\alpha\beta} \partial_{t_{n,\alpha}}
\partial_{t_{m-n,\beta}}Z \, .
\end{equation}
We still have to exhibit the exceptional constraints at $m=0$ or $m=1$, i.e. the puncture and dilaton equations. 
For $m=1$ there is an extra contribution from the  one-point function on the torus. Therefore, one has for $m=1$
\begin{equation}
    \partial_{t_1} Z= \sum_{n=0}^\infty \sum_{\alpha \in \Delta_m \cup \{ 0 \}} (c_{n,\alpha}+1) t_{n,\alpha} \partial_{t_{n,\alpha}}Z + \langle \sigma_1 \rangle_1 Z\, .
\end{equation}
For the puncture equation $m=0$, we obtain exceptional contributions when we recursively arrive at the three-point functions of primaries.  The pertinent terms in $\partial_{t_0} Z$ are $\frac{1}{2} \sum_{\alpha,\beta \in {\Delta}_m} t_{0,\alpha} t_{0,\beta} \langle P O_\alpha O_\beta \rangle $, and thus, one finds 
\begin{equation}
    \partial_{t_0} Z= \sum_{n=1}^\infty \sum_{\alpha \in\Delta_m \cup \{ 0 \}} (c_{n,\alpha}+1) t_{n,\alpha} \partial_{t_{n-1,\alpha}}Z + \frac{1}{2}\lambda^{-2}\sum_{\alpha,\beta \in {\Delta }_m} A_\alpha \eta_{\alpha\beta} t_{0,\alpha} t_{0,\beta} Z\, .
\end{equation}
To rewrite the constraints on the partition function in terms of a Virasoro algebra, we prepare the ground in several steps.
Firstly, we shift the time $t_{1,0} \to t_{1,0}+1$. Secondly, we relabel  $m \rightarrow m-1$ on the constraint equations.  Thirdly, we set
the times $t_{n,0}=0$, after the shift. The equations then take the form of Virasoro constraints
\begin{equation}
    L_{m \ge -1}Z=0 \, ,
\end{equation}
where the Virasoro generators are 
\begin{eqnarray}
L_{m \ge 1} &=& \sum_{n=0}^\infty \sum_{\alpha \in \Delta_m 
}(n+q_\alpha-\frac{c}{6}+\frac{1}{2}) t_{n,\alpha} \partial_{t_{m+n,\alpha}}+\frac{9-c}{6} \lambda^2 \sum_{n=1}^m \sum_{\alpha,\beta \in {\Delta }_m} b_\alpha \eta^{\alpha\beta} \partial_{t_{n-1,\alpha}}\partial_{t_{m-n,\beta}}
\nonumber\\
L_0 &=& \sum_{n=0}^\infty \sum_{\alpha \in \Delta_m 
}(n+q_\alpha-\frac{c}{6}+\frac{1}{2}) t_{n,\alpha} \partial_{t_{n,\alpha}}+ \frac{9-c}{6} \langle \sigma_1 \rangle_1
\nonumber\\
L_{-1} &=& \sum_{n=1}^\infty \sum_{\alpha \in \Delta_m 
}(n+q_\alpha-\frac{c}{6}+\frac{1}{2}) t_{n,\alpha} \partial_{t_{n-1,\alpha}}+\frac{9-c}{12}\lambda^{-2}\sum_{\alpha,\beta \in {\Delta }_m} A_\alpha \eta_{\alpha\beta} t_{0,\alpha}t_{0,\beta} \, . \label{Virasorogenerators}
\end{eqnarray}
These differential operators satisfy the (half) Virasoro algebra
\begin{equation}
    [L_m,L_n]=(m-n)L_{m+n}\, ,
\end{equation}
which implies that the constraints are mutually compatible. 
The consistency relation (\ref{consistency}) between the normalization $A_\alpha$, the parameters $b_\alpha$ and the one-point function $\langle \sigma_1 \rangle_1$ is crucial in closing the (half) Virasoro algebra.

Finally, let us stress that the shift in $t_{1,0}$ implies that the Virasoro constraints are on the partition function $F$ acted
upon by the exponential $\exp ( \int \sigma_{1,0})$. Only when the  operator $\sigma_{1,0}$ is in the spectrum can the shift be absorbed in the physical couplings of the theory.

\subsection{On Further Constraints}
Let us briefly comment on the extension of the Virasoro algebra constraints. For the compact model the Virasoro 
generators can be read as the energy momentum tensor modes for $k_c-1$ twisted bosons \cite{Li:1990fc}. For the non-compact model, this
is also the case. They correspond to the energy momentum tensor of $k-1$  twisted bosons.
The twist of the bosons is by the shifted R-charges $1/2-c/6+q_\alpha$ which again take values in the set
$\{ 1/k, 2/k, \dots , (k-1)/k \}$. 
The zero-point energy of the twisted bosons agrees with the zero point energy in the $L_0$ operator in (\ref{Virasorogenerators}).
The energy-momentum tensor is the energy-momentum tensor of a level one $A_{k-1}$  Kac-Moody algebra twisted by the Coxeter element of the Weyl group.

As in the compact model, one may suspect that the Virasoro algebra can  be  supplemented with higher spin currents to form a W-algebra. In order to analyze this conjecture in the conformal field theory framework, one needs a better grip on the contact terms among the operators $\sigma_{n,\alpha}$. Because of the special role of the unit operator in the non-compact models, this requires more work. Alternatively, one can characterize the W-algebra from the perspective of the Hamiltonian structure of the integrable hierarchy.
In any case, it would be interesting to further explore and exploit these constraints.

\section{The Critical Behaviour}
\label{scaling}
In this section, we exhibit the critical behavior of solutions to non-compact topological matter coupled to topological gravity.
We first determine the susceptibility and critical exponents by exploiting the dilaton and charge conservation constraints on the correlators.
Secondly, we confirm the result from the integrable hierarchy perspective. Finally, we  explore whether this new phase is related to proposals for critical exponents for $c>1$ matter systems coupled to ordinary two-dimensional gravity.

\subsection{The Scaling from the Differential Equations}

\label{scalingfromdiff}

In this subsection, we exploit the results of section \ref{recursion} to find the scaling behaviour of solutions. In particular, we use
the ghost number conservation equation (\ref{ghostnumberconservation}) as well as the dilaton constraint $L_0 Z=0$. The  conservation law
leads to the equation 
\begin{equation}
\sum_{n,\alpha} (n-1+q_\alpha) t_{n,\alpha} \partial_{n,\alpha} F_g = (g-1)(3-\frac{c}{3}) F_g \label{conservationequation} \, ,
\end{equation}
while the dilaton equation $L_0 Z=0$ can be rewritten 
\begin{equation}
\sum_{n,\alpha} (n+q_\alpha-\frac{c}{6}+\frac{1}{2}) t_{n,\alpha} \partial_{n,\alpha} F_g = \frac{9-c}{6} \langle \sigma_1 \rangle_1 \delta_{g,1} F_g \, , \label{dilatonequation}
\end{equation}
where we used the  conservation equation (\ref{conservationequation}).

To set the scale of the theory, we can turn on a coupling $t_{0,0}$ to the puncture operator $P=\sigma_{0,0}$. 
The parameter accompanying the puncture is intuitively argued to be related to the area because the puncture operator removes an area factor from
the invariance group of the surface \cite{Witten:1989ig}. 
Alternatively, to set the scale of the theory we can turn on the coupling to any other operator in the spectrum, and then determine the scaling behavior in terms of that coupling when turning on a second operator.
The intuitive  argument that another puncture coupling is also related to the area of the surface goes through. However, we can imagine the change to the dynamically gravitating surface to be more drastic when the puncture is accompanied by non-trivial matter.
The latter remark will be crucial when the puncture operator is not part of the spectrum
(because the identity operator is absent from the matter theory).  

After these preliminary remarks, we look for  scaling solutions.
We turn on a non-zero coupling $t_{0,\alpha_0}=x$ to a primary $\sigma_{0,\alpha_0}$ that plays the role of a generalized puncture operator, as argued above. We also turn on a second coupling $t_{n_0,\beta_0}=t$ to any other operator
$\sigma_{n_0,\beta_0}$ 
that will play the role of a generalized dilaton operator. Finally, we allow for a non-zero coupling $t_{0,\alpha}=t_\alpha$ to a primary field $\sigma_{0,\alpha}$. Using the equations  (\ref{conservationequation}) and (\ref{dilatonequation}), restricted to only these non-zero couplings, we find for $g \neq 1$:
\begin{align}
(q_{\alpha_0}-1) x \partial_x F_g + (n_0-1+q_{\beta_0})t\partial_t F_g 
+(q_\alpha-1) t_\alpha \partial_{t_\alpha} F_g
&= (g-1)(3-\frac{c}{3}) F_g \nonumber \\
(q_{\alpha_0}-\frac{c}{6}+\frac{1}{2}) x \partial_x F_g + (n_0+q_{\beta_0}-\frac{c}{6}+\frac{1}{2}) t \partial_t F_g
+(q_\alpha-\frac{c}{6}+\frac{1}{2}) t_\alpha \partial_{t_\alpha} F_g 
&=0\, .
\end{align}
Firstly, after setting $t_\alpha=0$, and eliminating the coupling $t$, we conclude that the scaling of the free energy $F_g$ at genus $g$ is  $F_g \propto  x^{2-\gamma_s}$
where the string susceptibility $\gamma_s$ at genus $g$ is 
\begin{equation}
\gamma_s(g,c,\alpha_0,\beta_0,n_0) = 2- (g-1) \frac{\frac{c}{3}-1-2 n_0 - 2 q_{\beta_0}}{n_0-q_{\alpha_0}+q_{\beta_0}} \, .  \label{genusrecursionscaling}
\end{equation}
At genus zero, this reduces to the susceptibility
\begin{equation}
\gamma_s = \frac{\frac{c}{3}-1-2 q_{\alpha_0}}{n_0+q_{\beta_0}-q_{\alpha_0}} \, . \label{recursionscaling}
\end{equation}
Secondly, we conclude from the scaling equations (after eliminating $t$ at non-zero $t_\alpha$) that the operators $\sigma_{0,\alpha}$  have  anomalous dimension 
\begin{equation}
\gamma(\sigma_{0,\alpha}) = \frac{q_\alpha-q_{\alpha_0}}{n_0+q_{\beta_0}-q_{\alpha_0}} \, .
\end{equation}
These are the critical exponents as a function of the R-charges, generalized puncture operator
$\alpha_0$ and dilaton operator $(n_0,q_{\beta_0})$.

\subsection{The Scaling from the Integrable Hierarchy}
\label{scalingfromintegrable}
In this subsection, we confirm the scaling behaviour computed in  subsection \ref{scalingfromdiff} at genus
zero by using the description of correlators in terms of the integrable model of sections \ref{noncompacthierarchy} and
\ref{noncompacthierarchywithgravity}. As before, we turn on the reference primary time $t_{0,\alpha_0}$ and  consider a solution with
non-zero $\tilde{t}_{N,\beta}=-\delta_{N,N_0} \delta_{\beta,\beta_0}$ (where $N_0 \ge 1$). Let us first sketch the re-derivation of the critical
behaviour in the case where the integrable model corresponds to a twisted, deformed compact $N=2$ superconformal field theory (where $k=0$). We first determine the scaling  dimension $[ L_c ]$ of the root $L_c$ when $\alpha_0,\beta_0 \ge 0$:
\begin{equation}
1 \equiv [ t_{\alpha_0} ] = [\langle  \sigma_{N_0-1} (\phi^{\beta_0}) \phi_{\alpha_0} \rangle ] = [ L_c^{N_0 k_c+\beta_0-\alpha_0} ]
\end{equation}
which implies
\begin{equation}
[L_c] = \frac{1}{N_0 k_c + \beta_0 - \alpha_0} \, .
\end{equation}
{From} this, we derive the dimension of the free energy 
\begin{equation}
[F] = [ t_{\alpha_0}^2 \langle (\phi^{\alpha_0})^2 \rangle] = 2 + \frac{2(\alpha_0+1)}{N_0 k_c + \beta_0 -\alpha_0}
\end{equation}
and the susceptibility\footnote{While our method of derivation is as proposed in \cite{Aoyama:1995kv}, the final result differs.}
\begin{equation}
\gamma_s = -2 \, \frac{\alpha_0+1}{N_0 k_c + \beta_0 - \alpha_0} \, \, . \label{compactsusceptibility}
\end{equation}
This matches the result (\ref{recursionscaling}) we obtained from the recursion relation and charge conservation.
The dimension of operators can similarly be checked.
When $\alpha_0 = 0$ in the compact model, we can rewrite the susceptibility as
\begin{equation}
\gamma_s =  (\frac{c}{3}-1) \frac{1}{N_0+q_{\beta_0}} \, ,
\end{equation}
which matches the analysis of \cite{Dijkgraaf:1990qw}. In particular, note that for $N_0+q_{\beta_0}=2$ we obtain
the susceptibility $\gamma_s=-1/k_c$. 
For the non-compact model, and assuming $\alpha_0,\beta_0 \le -2$, a similar derivation gives the susceptibility:
\begin{equation}
\gamma_s = 2 \frac{1 + \alpha_0}{ N_0 k - \beta_0 + \alpha_0} \, , \label{noncompactsusceptibility}
\end{equation}
which again agrees with the universal result (\ref{recursionscaling}).
Finally, for $\alpha_0 =0 $, $k_c=1$ and $\beta_0 \le -2$, we find from the integrable hierarchy:
\begin{equation}
\gamma_s = -\frac{2k }{ (N_0-1) k - \beta_0 - 1}    \label{intscaling}
\end{equation}
and we must recall that we took $N_0 \ge 1$.

\subsection{A Peculiar Phase of Gravity}
In this subsection, we evaluate to what extent  the coupling of non-compact matter to topological gravity leads to a new phase of gravity in two dimensions. Firstly, we succinctly review what is known, and then compare to the properties of our model, and how to interpret them.
 \subsubsection{Little Matter}
Two-dimensional quantum gravity coupled to matter was first solved using random matrix theory. 
The critical behavior was reproduced in the continuum approach to quantum gravity based on Liouville theory.
Moreover, a manifestly topological formulation of two-dimensional gravity was shown to also code the same critical exponents.
There is however a  restriction on the theories that were described in these three largely equivalent frameworks. The matter theories to which gravity is coupled cannot be too rich : they have central charge smaller or equal to one. 

One way to understand the barrier is in the continuum formulation, as follows.
The total central charge of the two-dimensional theory of gravity coupled to matter must equal zero, and moreover is equal to
\begin{equation}
c_{tot} = 0 = c_\phi + c_{gh} + c = 1+6 Q^2 -26 +c
\end{equation}
where the ghost central charge $c_{gh}$ in two-dimensional gravity is $c_{gh}=-26$,
the Liouville central charge $c_\phi$ equals $c_\phi=1+6Q^2$ and the matter central charge $c$ is initially
arbitrary.  However, the Liouville central charge $Q=b+b^{-1}$ is tied to the coefficient
$b$ in the Liouville potential $e^{b \phi}$. If the coefficient $b$ is real, than we have
that $c_\phi \ge 25$ and therefore that $c \le 1$, hence leading to a restriction on the amount of matter we can couple to two-dimensional gravity.  This  is the central charge barrier.\footnote{It is not useful to us to comment on the otherwise interesting possibility $c \ge 25$.}

These phases of gravity are characterized by critical exponents. These exponents measure the scaling of correlation functions of operators as a function of area. These are the gravitationally dressed dimensions of the operators. The susceptibility is a most basic critical exponent, measuring the scaling of the free energy with area. All these critical exponents can be obtained in the three approaches to two-dimensional quantum gravity. As an example we consider two-dimensional gravity coupled to a minimal model at central charge
\begin{equation}
c = 1 - \frac{6}{m(m+1)} \, 
\end{equation}
where $m$ is a positive integer. 
The susceptibility $\gamma_s$ of this theory at genus zero
is equal to 
\begin{equation}
\gamma_s(m) = -\frac{1}{m} \, .
\end{equation} 
For instance, pure gravity corresponds to susceptibility $\gamma_s=-1/2$. Clearly, there is a series of models of little matter coupled to gravity with susceptibilities that approach zero from below as $-1/m$, for $m$ going to plus infinity.

 \subsubsection{Positive Susceptibility with Little Matter}
 \label{Legendre}
 It is also interesting to us to mention a result that goes beyond this canon of two-dimensional gravity. In particular, we briefly recall theories of two-dimensional quantum gravity which give rise to positive susceptibility.
 Multi-trace deformations were added to matrix models and shown to give rise to positive susceptibilities 
 \cite{Das:1989fq,Korchemsky:1992tt,AlvarezGaume:1992np}. These can also be understood as minimal models with a dual gravitational dressing \cite{Klebanov:1994pv} which in the end gives rise to a new susceptibility
 \begin{equation}
 \bar{\gamma}_s = \frac{\gamma_s}{\gamma_s-1} \, .
 \end{equation}
 For unitary minimal models for example, the new susceptibility becomes
 \begin{equation}
 \bar{\gamma}_s = \frac{1}{m+1}
 \end{equation}
 which is positive, maximally $1/3$, and which tends to zero from above as the integer $m$ tends towards infinity. These models can be interpreted as lending a critical weight to wormholes, destroying the classical world sheet. They are still based on matter theories with central charge $c \le 1$. 
 These models are a Legendre transform \cite{Klebanov:1994kv} of the original models. These phenomena foreshadowed features of holography in anti-de Sitter space \cite{Aharony:2001pa,Witten:2001ua}. 
 
 \subsubsection{Positive Susceptibility with More Matter}
 There are speculations on how to couple  larger amounts of matter in two dimensions to gravity consistently, and to understand the resulting phase of quantum gravity. Over the years,  attempts have been made to calculate the consequences, and various scenarios have been envisaged. Lattice models of gravity coupled to matter systems with central charge $c>1$ (like multiple free bosons or  Ising spins) have been studied in numerical and analytic detail, with resulting predictions of a susceptibility  $\gamma_s=1/2$ at large matter central charge -- 
 the behavior of branched polymers --,
and an intermediate susceptibility  $\gamma_s=1/3$ at small central charge \cite{Ambjorn:1993ei,Ambjorn:1994ab,Durhuus:1994tu}. Analytic renormalization group flow methods have  attempted to explain these potentially universal dynamics \cite{David:1996vp}.
Thus, there are indications that $c>1$ gives rise to positive susceptibilities.

\subsubsection{A Fata Morgana}

Our models break the topological central charge barrier $c=3$. We can ask  whether they  give rise to positive susceptibilities. If we naively follow the universal conformal field theory formula (\ref{recursionscaling}), and put $\alpha_0=0=\beta_0$, $c=3+6/k$ and $N_0=2$, then we will find a  positive susceptibility
 \begin{equation}
 \gamma_s(k) \stackrel{?}{=} \frac{1}{k} \, , \label{tentative}
 \end{equation}
 where $k$ is a positive integer. It is tempting to read this as a continuum description of the tentative results for matter systems with $c>1$ coupled to ordinary gravity  reviewed above.
 
 However, this is too glib. The formula (\ref{recursionscaling}) is derived in a conformal field theory context, and in the non-compact conformal field theory at central charge $c=3+6/k$ the identity operator is non-normalizable. Therefore, a priori, the values $\alpha_0=0=\beta_0$ are disallowed and we cannot draw the positive susceptibility conclusion
 (\ref{tentative}). It is interesting to attempt to circumvent this conclusion. We can make a fair attempt by reverting to the integrable hierarchy framework, and studying a model at $k_c=1$ and level $k$ generic. As we have seen, in this model, we have added a unit operator (as well as the operator $\phi^{-1}$), and we can indeed pick $\alpha_0=0=\beta_0$. However,  the calculation of the susceptibility will proceed along different lines in the integrable model with extended spectrum, and gives a negative susceptibility (recorded in equation (\ref{compactsusceptibility})), while allowing $\beta_0 \le -2$ still gives rise to a negative susceptibility (see equation
 (\ref{intscaling})). Thus, we conclude that adding the identity operator to the non-compact conformal field theory in a manner consistent with integrability is incompatible with the law (\ref{tentative}). 
 
 The fata morgana (\ref{tentative}) is conceptually interesting. Non-compact matter punctures Riemann surfaces in a manner that is drastically different from compact matter, in the sense that non-compact matter punctures can apparently not be blindly analytically continued to compact matter punctures like the identity operator. Rather, after coupling to gravity, the coupled system mimics the behaviour of compact gravity e.g. in the fact that the generalized susceptibility (depending on $\alpha_0 \le -2$) is negative. The addition of more matter simultaneously changes the  gravitational backreaction 
and the combined system reverts to a familiar critical behaviour.

\section{Conclusions}
\label{conclusions}
We embedded the non-compact topological quantum field theory of \cite{Li:2018rcl} in the dispersionless Kadomtsev-Petviashvili integrable hierarchy. After coupling to gravity, the genus zero correlation functions are still governed by the  hierarchy, and we described how to compute all tree level amplitudes. Moreover, we generalized the Virasoro constraints to apply to the topological string with non-compact matter, and thus laid bare a structure that controls the perturbation theory at all genera. The scaling solutions of this phase of topological gravity were determined from these two perspectives, and we illustrated the peculiar nature of the phase of  non-compact matter coupled to gravity.

Clearly, more work is needed on various fronts.  Technically, one would like to exploit the dispersionful KP hierarchy and its tau-function to compute correlation functions at higher genera. 
Conceptually, it would be  interesting to clarify further to what extent going beyond the topological central charge barrier
 relates to ordinary gravity coupled to matter with central charge larger than one. The latter theory is expected to develop tachyons, or an instability towards the creation of macroscopic holes, while our topological gravity theory coupled to non-compact matter is manifestly well-defined. What is the relation between these phases of two-dimensional gravity  ? Furthermore, there is a challenge posed to the matrix model formulation of two-dimensional gravity to reproduce the amplitudes of non-compact topological gravity. In this context, we can mention a tentative connection between analytically continued matrix models and the cigar coset conformal field theory \cite{Witten:1991mk,Brezin:2012uc}.

We conclude that topological gravity coupled to non-compact matter opens up intriguing avenues for future research.

\section*{Acknowledgments}
It is a pleasure to thank our colleagues  for creating a stimulating research environment. We
acknowledge support from the grant ANR-13-BS05-0001.

\appendix

\section{Concrete Examples}
\label{examples}
In this appendix, we give concrete
examples of (generalized) generating functions \cite{Dijkgraaf:1990qw} for correlators for non-compact topological quantum field theories, and non-compact topological gravity.

\subsection{A  Field Theory Hierarchy and the Strict Limit}
\label{TQFTexample}
We wish to give an example generating function $F_m$ computed in the formalism of \cite{Aoyama:1995kv}, and concretely illustrate how to take the limit to the solution of the non-compact topological field theory calculated in \cite{Li:2018rcl}.
We have that the universal coordinate $u_{k_c-1}$ equals the pole location $s=u_{k_c-1}$ and that the  coordinate $u_{k_c-1}$ is conjugate, by the topological metric $\eta_{\alpha \beta}$ to the universal coordinate
$u_{-1}=v_{-1}$. Moreover, we have that the coordinate $u_0$ is conjugate to $u_{k_c-2}$. We want to pick the smallest allowed level $k_c$
in the approach of \cite{Aoyama:1995kv} in order to be as close to our target theory as possible and therefore choose $k_c=1$.
In this case, we can either parameterize $u_0$ as the pole, or as the constant term in the potential. Our goal is to eliminate the linear, constant and $X^{-1}$ term in the superpotential in order to find the non-compact superpotential of \cite{Li:2018rcl} which starts with a $X^{-2}$ term.
We have described how to take this limit on the superpotential in subsection \ref{limit}. We will now follow how this limit operates on a larger set of formulas from \cite{Aoyama:1995kv} and how the limit indeed allows us to recuperate the solution of \cite{Li:2018rcl}, all the way to the generating function. We work with the example of $k=3$.

The detailed formulas for $k_c=1$ and $k=3$, following \cite{Aoyama:1995kv} start from the superpotential
$W$
\begin{equation}
W = X  + \frac{v_{-1}}{X-s}
+ \frac{v_{-2}}{2(X-s)^2}+\frac{v_{-1}}{3 (X-s)^3} \, .
\end{equation}
We shift the pole $s$ to zero
\begin{equation}
W = X +s+ \frac{v_{-1}}{X}
+ \frac{v_{-2}}{2 X^2}+\frac{v_{-1}}{3 X^3} \, .
\end{equation}
The universal coordinates and the coefficients in the potential are related by
\begin{equation}
s = u_0 \, , \quad 
v_{-1} = u_{-1}\, , \quad
v_{-2} = 2 u_{-2} u_{-3} \, , \quad
v_{-3} = u_{-3}^3 \, .
\end{equation}
We find the superpotential and operators
\begin{eqnarray}
& & W = X+ u_0 + \frac{u_{-1}}{X} + \frac{u_{-2} u_{-3}}{X^2}+
\frac{u_{-3}^3}{3 X^3}  \, .
\nonumber \\
& & \phi^0 = \partial_{u_0} W = 1 \, , \quad
    \phi^{-1} = \partial_{u_{-1}} W = 1/X \, ,
    \nonumber \\
& & \phi^{-2} = \partial_{u_{-2}} W = u_{-3}/X^2 \, , \quad
 \phi^{-3} = \partial_{u_{-3}} W = u_{-2} X^{-2}+  u_{-3}^2 X^{-3} \, .
\end{eqnarray}
After some elementary  calculations, e.g. through the calculation of the second derivatives
$G^{\alpha \beta}$, the potential $F_m$ is found to be
\begin{equation}
F_m= -\frac{u_{-2}^4}{12 u_{-3}^2} +\frac{1}{2}  u_{-1}^2 \log u_{-3} + \frac{1}{2} u_{-1} u_0^2 + 
u_0 u_{-2} u_{-3}  + \frac{1}{2} \frac{u_{-1} u_{-2}^2 }{u_{-3}} 
+ \frac{1}{6} u_{-3}^3 \, .
\end{equation}
Let's consider  the limit 
 $X \rightarrow \epsilon X$ as well as 
$u_{-i \le -1} \rightarrow \epsilon u_{-i \le -1}$ while keeping $u_0$ fixed for now. We then obtain the potential
\begin{eqnarray}
W &=& \epsilon X +  u_0 + \frac{u_{-1}}{X} + \frac{u_{-2} u_{-3}}{X^2}+
\frac{u_{-3}^3}{3 X^3}  \, .
\end{eqnarray}
The scaling of the generating function $F_m$ depends on the term under consideration:
\begin{equation}
F_m(\epsilon) = \epsilon \frac{u_{-1} u_0^2}{2} + \epsilon^2 (
 -\frac{u_{-2}^4}{12 u_{-3}^2} +\frac{1}{2}  u_{-1}^2 \log (\epsilon u_{-3}) + 
u_0 u_{-2} u_{-3} + \frac{1}{2} \frac{u_{-1} u_{-2}^2}{u_{-3}} ) + \epsilon^3 \frac{u_{-3}^3}{6}  \, .
\end{equation}
We wish to put $u_{-1}=0=u_0$, and keep the leading terms in $\epsilon$. This reproduces
the solution 
\begin{equation}
\frac{F_m(\epsilon)}{\epsilon^2} = -\frac{u_{-2}^4}{12 u_{-3}^2}
\end{equation}
of the non-compact model at level $3$ obtained in \cite{Li:2018rcl}. It is not hard to check that one can also implement
the limit on the formulas
of \cite{Aoyama:1995kv} from the get-go, and find agreement once more.

\subsection{A Non-Compact Topological Gravity}
 \label{gravitationalexample}
In this appendix, we discuss a non-compact matter model coupled to gravity, with a non-trivial coupling to gravity turned on. We again pick the model to have level $k=3$, and study only the model after taking the strict limit, such that we have primaries $\alpha \in \{ -2,-3 \}$. The pure matter model is described in appendix \ref{TQFTexample}. To describe the matter model coupled to gravity, we start out by stating
explicitly the constitutive relations
\begin{eqnarray}
G^{-2,-2}=\langle \phi^{-2} \phi^{-2} \rangle &=& - \frac{u_{-2}^2}{u_{-3}^2} =\tilde{u}^{-2}
\nonumber \\
G^{-2,-3}=\langle \phi^{-2} \phi^{-3} \rangle &=& \frac{2}{3} \frac{u_{-2}^3}{u_{-3}^3}
= \tilde{u}^{-3}
\nonumber \\
G^{-3,-3}=\langle \phi^{-3} \phi^{-3} \rangle &=& -\frac{1}{2} \frac{u_{-2}^4}{u_{-3}^4}
=-\frac{1}{2} (\tilde{u}^{-2})^2 \, .
\end{eqnarray}
The variables $\tilde{u}^\alpha$ are the universal coordinates for the choice $\alpha_0=-2$. We only allow for non-zero primary times
$t_{-2}=t_{-2,0}$ and $t_{-3}=t_{-3,0}$ as well as one gravitational descendant time $t_{1,-2}$.
We choose  the shifts  $C_{N,\alpha}=- \delta_{N,1} \delta_{\alpha,-2}$ 
which lead to the following renormalized relation between the universal coordinates $\tilde{u}_\alpha$ and the time variables
\begin{eqnarray}
\tilde{u}_{-2} &=& \hat{t}_{-2} = t_{-2} + \tilde{u}^{-3} t_{1,-2}
\nonumber \\
\tilde{u}_{-3} &=& \hat{t}_{-3} = t_{-3} + \tilde{u}^{-2} t_{1,-2} \, .
\end{eqnarray}
We also have 
\begin{eqnarray}
\tilde{t}_{N,\alpha} &=& t_{N,\alpha} - \delta_{N,1} \delta_{\alpha,-2} \, .
\end{eqnarray}With the use of the  two-point function of descendants 
\begin{eqnarray}
\langle \sigma_1(\phi^{-2}) \phi^{-2} \rangle &=& \langle \phi^{-5} \phi^{-2} \rangle =
- \frac{2}{3} \frac{u_{-2}^5}{u_{-3}^5} = 
\frac{2}{3}  (-\tilde{u}^{-2})^{\frac{5}{2}}
\nonumber \\
\langle \sigma_1(\phi^{-2}) \phi^{-3} \rangle &=& \langle \phi^{-5} \phi^{-3} \rangle = 
\frac{5}{9} \frac{u_{-2}^6}{u_{-3}^6} = - \frac{5}{9} (\tilde{u}^{-2})^3
\nonumber \\
\langle \sigma_1 (\phi^{-2}) \sigma_1 (\phi^{-2})\rangle  &=& \langle \phi^{-5} \phi^{-5} \rangle
= -\frac{25}{36} \frac{u_{-2}^8}{u_{-3}^8} =-\frac{25}{36} (\tilde{u}^{-2})^4 \, ,
\end{eqnarray}
we can then  compute the generating function of  topological gravity correlation functions
\begin{eqnarray}
F_0 
&=& \frac{1}{2} t_{-2}^2 \tilde{u}^{-2}
+ t_{-2} t_{-3} \tilde{u}^{-3} 
- \frac{1}{4} t_{-3}^2  (\tilde{u}^{-2})^2
\nonumber \\
& & 
+ (t_{1,-2}-1) t_{-2}  \frac{2}{3}  (-\tilde{u}^{-2})^{\frac{5}{2}}
- (t_{1,-2}-1) t_{-3}  \frac{5}{9} (\tilde{u}^{-2})^3
\nonumber \\
& & 
-  (t_{1,-2}-1)^2 \frac{25}{72} (\tilde{u}^{-2})^4 \, .
\end{eqnarray}
One can express the generating function in terms of the time variables only.
This provides one example of a generating function in the presence of a non-zero coupling
for a gravitational descendant.

\bibliographystyle{JHEP}

\end{document}